\documentclass[aps,prl,reprint,superscriptaddress,preprintnumbers]{revtex4-1}
 \usepackage{graphicx}
 
\begin{document}

\title{High resolution coherent population trapping on a single hole spin in a semiconductor}

\author{Julien Houel}
\affiliation{Department of Physics, University of Basel, Klingelbergstrasse 82, CH4056 Basel, Switzerland}

\author{Jonathan H. Prechtel}
\email[]{jonathan.prechtel@unibas.ch}
\homepage[]{http://nano-photonics.unibas.ch/}

\affiliation{Department of Physics, University of Basel, Klingelbergstrasse 82, CH4056 Basel, Switzerland}

\author{Daniel Brunner}
\affiliation{Instituto de Fisica Interdisciplinar y Sistemas Complejos, IFISC (UIB-CSIC), Campus Universitat de les Illes Balears, Palma de Mallorca E-07122, Spain}

\author{Christopher E. Kuklewicz}
\affiliation{SUPA, Institute of Photonics and Quantum Sciences, Heriot-Watt University, Edinburgh, United Kingdom}

\author{Brian D. Gerardot}
\affiliation{SUPA, Institute of Photonics and Quantum Sciences, Heriot-Watt University, Edinburgh, United Kingdom}

\author{Nick G.\ Stoltz}
\affiliation{Materials Department, University of California, Santa
Barbara, California 93106, USA}

\author{Pierre M.\ Petroff}
\affiliation{Materials Department, University of California, Santa
Barbara, California 93106, USA}

\author{Richard J. Warburton}
\affiliation{Department of Physics, University of Basel, Klingelbergstrasse 82, CH4056 Basel, Switzerland}

\date{\today}

\begin{abstract}
We report high resolution coherent population trapping on a single hole spin in a semiconductor quantum dot. The absorption dip signifying the formation of a dark state exhibits an atomic physics-like dip width of just 10 MHz. We observe fluctuations in the absolute frequency of the absorption dip, evidence of very slow spin dephasing. We identify this process as charge noise by, first, demonstrating that the hole spin g-factor in this configuration (in-plane magnetic field) is strongly dependent on the vertical electric field, and second, by characterizing the charge noise through its effects on the optical transition frequency. An important conclusion is that charge noise is an important hole spin dephasing process.
\end{abstract}

\maketitle

Coherent population trapping (CPT) is a quantum interference effect which arises in an optical $\Lambda$ system \cite{Arimondo1996}. Two ground states are coupled individually by ``pump" and ``probe" lasers to a common upper level. At the two-photon resonance, that is, when the frequency difference of the lasers matches the frequency difference of the ground states, one of the three eigenstates has zero amplitude of the upper level, the ``dark" state. This zero amplitude can be thought of as a quantum interference between the two excitation pathways. CPT refers to the signature of the dark state, a dip in the probe absorption spectrum as the probe is tuned through the two-photon resonance. Specifically, for probe and pump couplings $\hbar \Omega_{1}$, $\hbar \Omega_{2}$ in the perturbative regime $\hbar \Omega_{1} \ll \hbar \Omega_{2} \ll \hbar \Gamma_{r}$ ($\Gamma_{r}$ is the spontaneous emission rate from the upper state), the dip has width $\hbar \Omega_{\rm 2}^{2}/\Gamma_{r}$.

CPT is a key effect in atomic physics. First, CPT forms the microscopic origin of electromagnetically-induced transparency which itself underpins a scheme for slowing light \cite{Fleischhauer2005}. Extremely small light group velocities have been demonstrated \cite{Hau1999}. Secondly, the two-photon resonance has spectroscopic implications: the narrow dip enables the frequency separation of the ground states to be measured extremely precisely by optical means. Thirdly, the dark state of CPT can be used for quantum control. For instance, by adiabatically switching one laser on as the other is turned off, the population can be transferred from one ground state to the other without ever occupying the upper state \cite{Bergmann1998}. Finally, the ``visibility" of the quantum interference at the CPT dip is sensitive to the ground state coherence (but insensitive to the upper state coherence) \cite{Imamoglu2006,Xu2008,Brunner2009,Weiss2012}: spectroscopic characterization of the dip is a powerful way of measuring the decoherence and dephasing times, a complementary technique to a Ramsey fringe/Hahn echo measurement. The sensitivity to decoherence comes about because ground state decoherence admixes the dark state with the two bright states. The dip width sets the sensitivity of the experiment to the ground state decoherence rate $\gamma$: for $\gamma \ll \Omega_{\rm 2}^{2}/\Gamma_{r}$ the signal in the dip goes to zero (high ``visibility") but for $\gamma \gg \Omega_{2}^{2}/\Gamma_{r}$ the dip is washed out (low ``visibility").

It is clearly motivating to implement CPT in a semiconductor with the perspective of using a solid-state system as the host for slow light, quantum  metrology and quantum control. In addition, CPT offers a spectroscopic means of exploring the complex decoherence and dephasing mechanisms in the solid-state. Optical semiconductor-based CPT experiments to date include excitons in GaAs quantum wells \cite{Phillips2003}, bound excitons in GaAs \cite{Sladkov2010}, an electron spin bound to an InGaAs quantum dot \cite{Xu2008}, a hole spin bound to an InGaAs quantum dot \cite{Brunner2009}, and the two-electron state of a quantum dot molecule \cite{Weiss2012}. In all these experiments, the key challenge is to engineer two ground states with decoherence and dephasing times ($T_{2}$ and $T_{2}^{*}$, respectively) much larger than the radiative lifetime $\tau_{r}$ of the upper state. An InGaAs quantum dot is an obvious candidate: a single electron can be trapped in the quantum dot, spin providing a natural two-level system, and the upper level decays quickly by spontaneous emission, $\tau_{r} \sim 1$ ns \cite{Dalgarno2008a}. Unfortunately, in the presence of noisy nuclei, the electron spin-nuclear spin hyperfine interaction limits $T_{2}^{*}$ to just a few ns \cite{Merkulov2002,Khaetskii2002} and the CPT dip can only be observed at large optical couplings where it is inevitably broad \cite{Xu2008}. The situation improves either by reducing the nuclear spin noise \cite{Xu2009} or by using a quantum dot molecule \cite{Weiss2012} at an atomic clock-like point where the first order sensitivity to spin noise vanishes. 

A hole spin is potentially simpler. The motivation is that a close-to-ideal heavy hole state with spin eigenvectors $J=\frac{3}{2}, J_{z}=\pm\frac{3}{2}$ is predicted to become coherent in an in-plane magnetic field \cite{Fischer2008}. Conveniently, the in-plane magnetic field is exactly the field direction required to establish the $\Lambda$ system \cite{Xu2008,Brunner2009}. The key point is that a perfect heavy hole spin is sensitive to nuclear spin noise only along the vertical direction $z$, noise which can be suppressed by applying an external magnetic field in the $(x,y)$-plane \cite{Fischer2008}. The extent to which the idealized heavy hole picture applies to a real hole in a quantum dot has been explored in a number of recent optical experiments \cite{Gerardot2008,Brunner2009,Heiss2008,Ramsay2008,Godden2010,Crooker2010,DeGreve2011,Greilich2011}. The Hahn echo $T_{2}$ is in the $\mu$s range \cite{DeGreve2011}, a remarkable result bearing in mind the extremely limited hole spin coherence in quantum wells and bulk material \cite{Damen1991,Marie1999}. Spin dephasing times $T_{2}^{*}$ lie in the $10-100$ ns regime with significant differences from experiment to experiment \cite{Brunner2009,Crooker2010,DeGreve2011,Greilich2011}. While CPT dips have been observed on a single hole spin \cite{Brunner2009}, these experiments used optical couplings only slightly less than the radiative decay rate, resulting in dip widths of $\sim 100$ MHz. This is too large for slow light, high resolution spectroscopy and quantum metrology applications. It also renders the experiment insensitive to decoherence times above about 100 ns.

We report here CPT on a single hole spin in the perturbative regime. A dip width of just 10 MHz is demonstrated. The residual absorption in the center of the dip is zero (more precisely, smaller than our experimental resolution of a few \%) such that the dark state is at most weakly admixed with the bright states, consistent with a coherence time $T_{2} \ge 1$ $\mu$s. However, we discover a scan-to-scan variation in the CPT position, an effect which was obscured by the larger dip widths in previous experiments. We relate this to charge noise. On the one hand, we quantify the dependence of the hole spin g-factor on vertical electric field. On the other hand, we characterize the fluctuations in vertical electric field through their effects on the optical transition, the dependence arising via the dc Stark effect. This allows us to identify charge noise as an important dephasing mechanism for the quantum dot hole spin.

\begin{figure}[t]
\vspace{0.0cm}                                 
\centering                                     
\includegraphics[width=\linewidth]{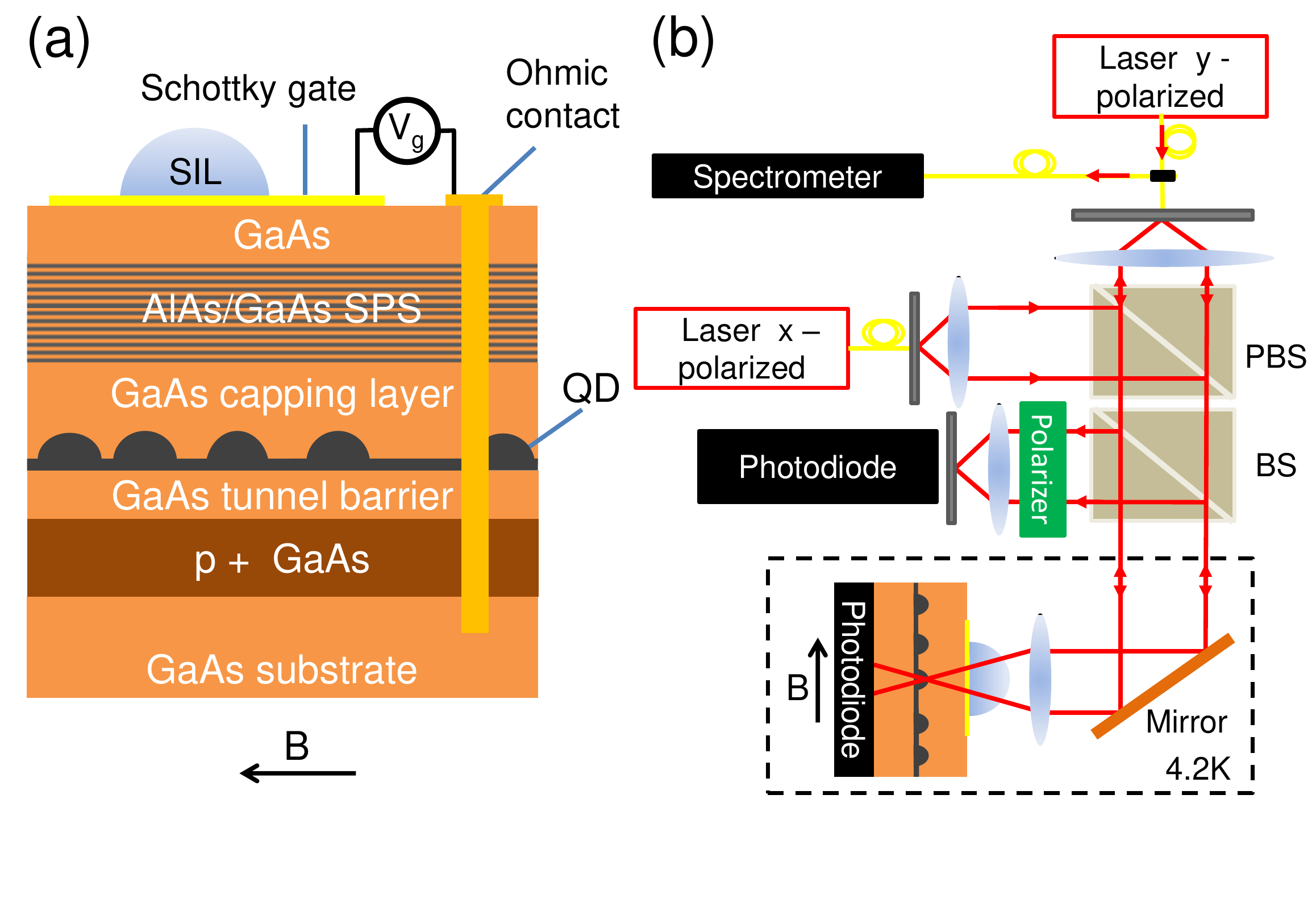}   
\vspace{-0.8cm} 
\caption[]{(Color online)
(a) Layer sequence of the semiconductor heterostructure. The distance back contact to quantum dot layer (tunnel barrier) is 25 nm, quantum dot layer to short-period superlattice (SPS) 10 nm (capping layer), SPS 120 nm. Post-growth, Ohmic contacts along with a semi-transparent surface gate electrode are fabricated, and a high-index glass hemispherical solid-immersion lens (SIL) is positioned on the sample surface. (b) Schematic of the optical set-up.}  
\label{figure1}                                       
\end{figure}

The semiconductor device is a p-type charge-tunable heterostructure consisting of a p$^{+}$ back contact (C doping), a 25 nm tunnel barrier, a layer of InGaAs quantum dots (optical wavelengths around 950 nm), a 10 nm GaAs capping layer, a 120 nm AlAs/GaAs superlattice blocking barrier, and a semi-transparent gate electrode to which a voltage $V_{g}$ is applied, Fig.\ 1(a) \cite{Gerardot2008,Brunner2009,Gerardot2011}. The $V_{g}$ regime in which one hole is trapped in the quantum dot is located and laser spectroscopy is carried out using differential reflectivity ($\Delta R/R$) detection with a room temperature p-i-n photodiode, Fig.\ 1(b) \cite{Hogele2004,Alen2006}. The exciton's optical linewidth is dot dependent, typically $5-10$ $\mu$eV \cite{Gerardot2011}. This is considerably larger than both the transform limit, $\sim 0.8$ $\mu$eV \cite{Kuhlmann2013}, and the linewidths on high quality n-type samples, $\sim 1.5$ $\mu$eV \cite{Hogele2004,Kuhlmann2013}, and reflects additional charge noise associated with the p-type doping \cite{Gerardot2011}. A magnetic field of 0.5 T is applied in the plane, and then two-color pump-probe laser spectroscopy is carried out using two coherent lasers (each with linewidth $1-2$ MHz measured in 1 s). The experiment is very challenging as, first, $\Delta R/R$ is very small at the optical resonance (0.1\%); second, at the ultra-low laser powers used here noise in the detector circuit is significant; and third, in the perturbative regime ($\hbar \Omega_{1} \ll \hbar \Omega_{2} \ll \hbar/\tau_{r}$), the width of the CPT dip approaches the limit set by the mutual coherence of the lasers ($\sim 10-20$ MHz measured in one minute). We meet these challenges with a glass solid immersion lens (refractive index = 2.0) to reduce the spot size in order to boost the $\Delta R/R$ signal, a modulation technique to reject noise in the reflectivity signal, and integration times of 10 s per point. We accept the wanderings in the pump laser frequency and a stabilization scheme locks the pump-probe frequency difference to a radio frequency reference resulting in a measured mutual coherence of 2.0 MHz (integration  30 s). 

A CPT dip on a single hole spin in a magnetic field of 0.5 T and temperature 4.2 K is shown in Fig.\ 2. The optical couplings $\hbar \Omega_{1}$, $\hbar \Omega_{2}$ were determined by measuring an Autler-Townes splitting at high laser powers, extrapolating the couplings to low laser powers using the scaling $\hbar \Omega \propto \sqrt{P}$ ($P$ is the laser power), Fig.\ 3(a). In the CPT experiment, $\hbar \Omega_{2}$ is a factor of 3 lower than the spontaneous decay rate $\Gamma_{r}=\hbar/\tau_{r}$. The full-width-at-half-maximum of the CPT dip is just 13 MHz, equivalently 54 neV. This width corresponds to just $10^{-7}$ of the frequency of the optical transitions, and just $10^{-3}$ of the thermal energy. Fig.\ 2 constitutes our main result: observation of a CPT dip linewidth in the MHz regime, a spectral sensitivity usually associated with atomic physics and not a semiconductor experiment.

\begin{figure}[t]
\vspace{0.0cm}                                 
\centering                                     
\includegraphics[width=\linewidth]{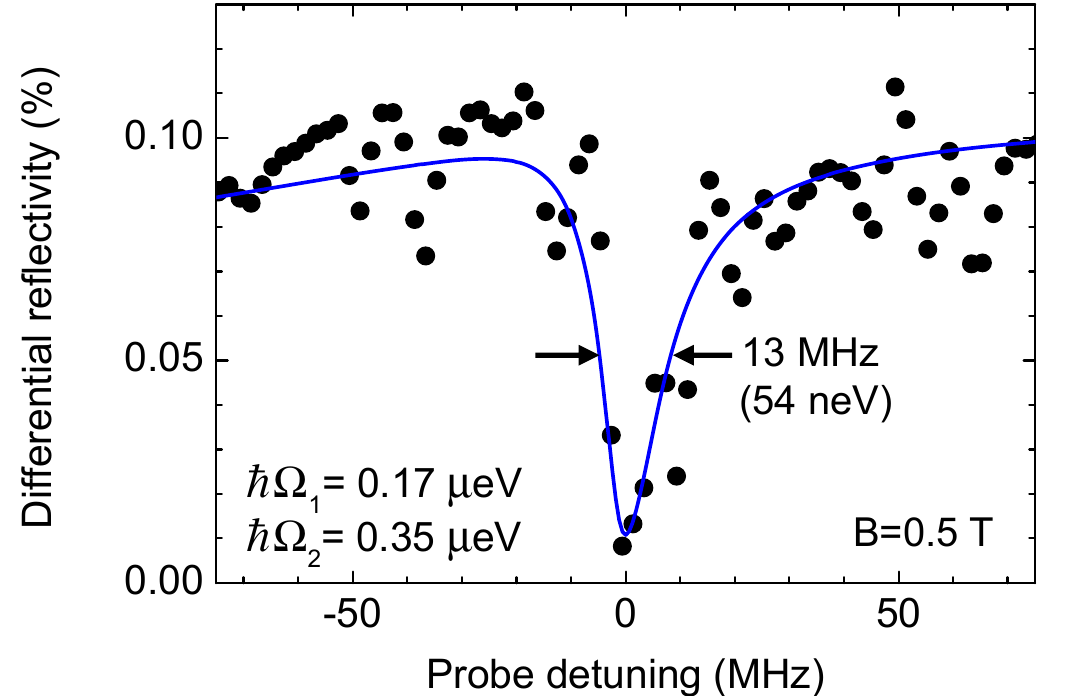}   
\vspace{-0.5cm} 
\caption[]{(Color online)
Probe absorption in the presence of a close-to-resonant pump laser on a single InGaAs quantum dot containing a single hole spin at $B=0.5$ T and $T=4.2$ K, 10 s integration per point. The absorption (here, differential reflectivity $\Delta R/R$) shows an absorption dip signifying coherent population trapping. The solid line shows the result of a 3-level density matrix model. The atomic physics result ($\hbar \Omega_{\rm 1}=0.17$, $\hbar \Omega_{\rm 2}=0.35$, $\hbar \delta_{2}=-2.5$ $\mu$eV, $\hbar/\tau_{r}=0.8$ $\mu$eV, $T_{2}\ge 1$ $\mu$s, $T_{1} \gg T_{2}$) is convoluted with a Lorentzian with FWHM $\Gamma_{X}=5$ $\mu$eV (1.22 GHz) to describe slow exciton dephasing, and then with a Lorentzian with FWHM 8.3 neV (2.0 MHz) to account for the mutual coherence of the lasers.}  
\label{figure2}                                       
\end{figure}

The solid-lines in Fig.s 2 and 3(a) correspond to the results of a 3-level density matrix ($\rho$) calculation \cite{Brunner2009}. The 3-level atomic physics result for $\rho_{13}$ \cite{Fleischhauer2005} is convoluted with a Lorentzian distribution of width $\Gamma_{X}$ for the energy of the upper level, $E_{3}$, in order to describe the effects of charge noise. $\Gamma_{X}$ is known from the one-laser characterization at $B=0$. The result is then convoluted again, this time with a Lorentzian function of width 2.0 MHz (8.3 neV) to describe the limited mutual coherence of the lasers. In the limit of large $\hbar \Omega_{2}$, the Autler-Townes experiment, $\hbar \Omega_{2}$ is treated as a fit parameter, and the result describes the absorption envelope extremely well, Fig.\ 3(a). In the limit of small $\hbar \Omega_{2}$, the CPT experiment, the result describes the dip width and depth extremely well (Fig.\ 2). In the CPT limit, there are no unknowns apart from a small uncertainty in $\hbar \delta_{2}$, the pump-detuning (see below). Even in this low-coupling limit, the residual signal in the CPT dip can be fully accounted for by the mutual coherence of the lasers and not by ground state decoherence, and the dip width is described by the 3-level atom and not by ground state dephasing. 

\begin{figure}[b]
\vspace{0.0cm}                                 
\centering                                     
\includegraphics[width=\linewidth]{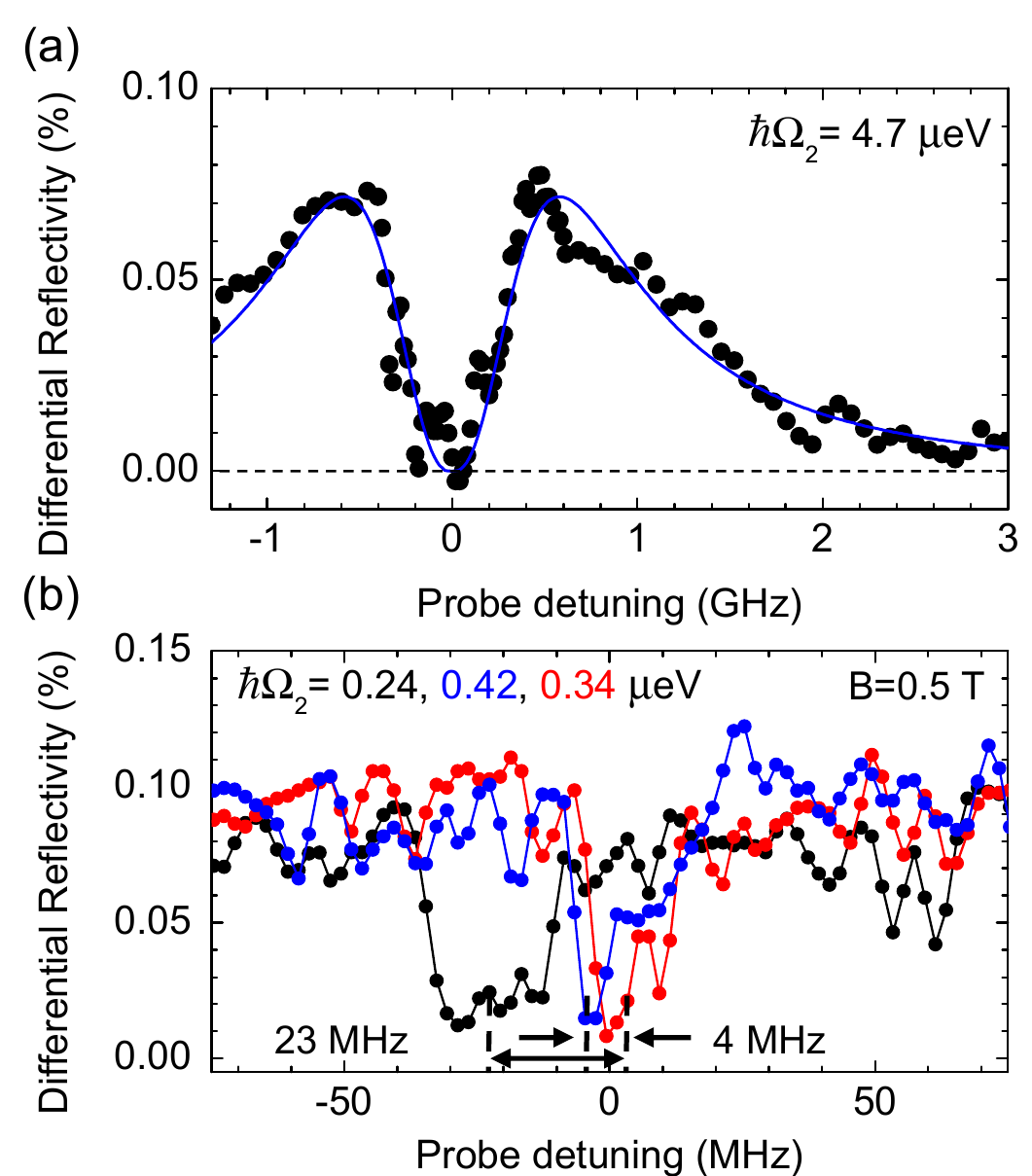}   
\vspace{-0.5cm} 
\caption[]{(Color online)
(a) Probe absorption versus probe detuning on the same quantum dot as Fig.\ 2 in the regime $\hbar \Omega_{2} \gg \hbar/\tau_{r}$ showing an Autler-Townes splitting. The solid curve is a fit to the data, $\hbar \Omega_{2}=4.7$ $\mu$eV, $\hbar \delta_{2}=0.0$ $\mu$eV, as fit parameters. $\Gamma_{X}=7.5$ $\mu$eV is taken from $\Delta R/R$ at $B=0$.
(b) Probe absorption versus probe detuning on the same quantum dot in the regime $\hbar \Omega_{2} \ll \hbar/\tau_{r}$ showing a CPT dip, as in Fig.\ 2. Three curves are shown under close-to-identical conditions showing a shift in the location of the dip in one case by 4 MHz (17 neV), and in another case by 23 MHz (95 neV). }  
\label{figure3}                                       
\end{figure}

We take advantage of the narrow dip to perform high resolution dark state spectroscopy and probe the limits of hole spin dephasing in our sample. We find that the frequency of the dip, i.e.\ the two photon resonance, fluctuates from scan to scan, see Fig.\ 3(b). The fluctuations occur in a frequency range of $\sim 5$ MHz ($\sim 20$ neV). Occasionally, larger frequency shifts are observed, possibly with an unusual lineshape. An example is shown in Fig.\ 3(b). These effects point to the presence of slow fluctuations in the frequency separation of the hole spin ground states. It is highly unlikely that nuclear spin noise is responsible. On the one hand, nuclear spin noise lies at higher frequencies \cite{Kuhlmann2013}; on the other hand, we have not observed any hysteresis effects in the CPT experiment, the typical signature of a coupling to the nuclear spins. The slow CPT fluctuations are reminiscent of the low frequency wanderings of the optical transition which has a $1/f$-like noise spectrum \cite{Kuhlmann2013}. Significantly, this flicker noise has its largest noise powers at the lowest frequencies. Its origin is however charge noise which results in a noisy electric field at the location of the quantum dot. We therefore look for a link between charge noise and the hole g-factor.

We have characterized the eigenenergies as a function of $V_{g}$. Specifically, on a quantum dot in the same device we measure the photoluminescence (PL) from the positively-charged trion X$^{1+}$ in an in-plane magnetic field $B=9$ T, Fig.\ 4(a). We resolve 4 lines corresponding to the two ``vertical" transitions and the two ``diagonal" transitions, Fig.\ 4(a),(b). With the assumption that the electron spin g-factor $g_{e}$ is negative, there is sufficient information to determine $g_{e}$ and $g_{h}$. Fig.\ 4 shows the $V_{g}$ dependence of the electron and hole Zeeman energies. To within error, $\sim 0.25$ \%, the electron Zeeman energy $E_{Z}^{e}$ is independent of $V_{g}$; in contrast, the hole Zeeman energy $E_{Z}^{h}$ changes by $\sim 5$\% over the X$^{1+}$ plateau. Defining the $g_{h}$ via $E_{Z}^{h}=g_{h} \mu_{B} B$, we find $g_{h}=0.15 + \alpha F$ with $\alpha=8.6 \times 10^{-4}$ cm/kV. The origin of the electric field dependence of the in-plane hole g-factor is probably related to the strong dependence of the bulk g-factor on material composition: as the field changes, the hole wave function experiences a different material environment with a concomitant effect on the g-factor. 

\begin{figure}[t]
\vspace{0.0cm}                                 
\centering                                     
\includegraphics[width=\linewidth]{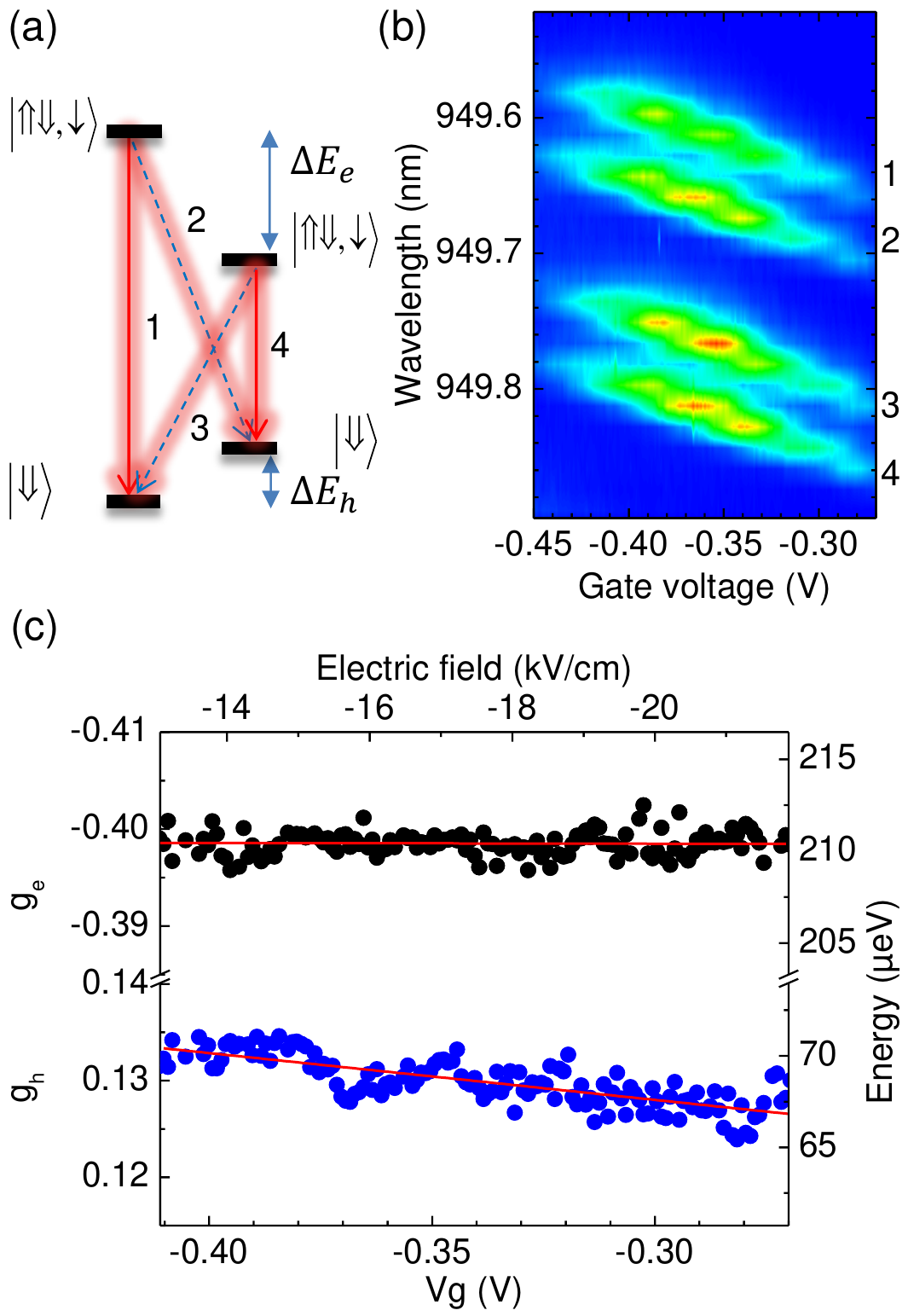}   
\vspace{-0.5cm} 
\caption[]{(Color online)
(a) The quantum states of a single hole spin in an in-plane magnetic field. $\uparrow,\downarrow$ denotes an electron spin, $\Uparrow,\Downarrow$ a hole spin. (b) Photoluminescence on a single InGaAs quantum dot (different quantum dot from Fig.\ 2, 3 but from the same wafer) at $B=9.0$ T and $T=4.2$ K as a function of $V_{g}$ over the extent of the single hole Coulomb blockade plateau. Four transitions are visible, labeled $1-4$, and identified in (a). The width of each peak is determined by the spectrometer-detector (system resolution 50 $\mu$eV). (c) Electron and hole g-factors, $g_{e}$ and $g_{h}$, versus $V_{g}$. (The hole Zeeman energy is defined as $E_{Z}^{h}=g_{h} \mu_{B}B$.) $V_{g}$ is converted into vertical electric field $F$ with $F=-e(V_{g}+V_{o})/D$ with $D=155$ nm and Schottky barrier $V_{o}=0.62$ V. Under the assumption of a negative $g_{e}$, $g_{h}$ is positive.}
\label{figure4}                                       
\end{figure}

The dependence of $g_{h}$ on $F$ creates a mechanism by which charge noise can result in spin dephasing: electric field fluctuations cause changes to the hole spin precession frequency. In particular, the low frequency charge noise causes both the optical transition energy and the CPT dip position to wander. Specifically, in these experiments, the charge noise in a bandwidth $f_{\rm scan} \le f \le 1/\tau_{r}$ can be determined by measuring the optical linewidth at rate $f_{\rm scan}$, converting the inhomogeneous broadening into electric field noise via the known Stark shift. The charge noise broadens the optical resonance above the transform limit, an effect which is easy to measure with the resonant laser spectroscopy employed here. Charge noise at lower frequencies, $f \le f_{\rm scan}$, results in scan-to-scan fluctuations of the resonance energy. CPT is recorded on the same quantum dot experiencing the same noise: a powerful connection can therefore be made from the optical response to the CPT experiment. 

The optical linewidth of the particular quantum dot in Fig.\ 3 is 7.5 $\mu$eV for $f_{\rm scan}=0.1$ Hz. Together with the known Stark shifts, $dE/dF=18$ $\mu$eV/kVcm$^{-1}$ [$dE/dV_{g}=1.12$ $\mu$eV/mV], this results in $\Delta F =0.42$ kV/cm [$\Delta V_{g}=6.7$ mV]. This charge noise, $\Delta F$, results in turn in a fluctuation in $g_{h}$ of $\Delta g_{h}=3.6 \times 10^{-4}$, equivalently, $\Delta E_{Z}^{h}=11$ neV (2.5 MHz) at 0.5 T. This broadening is comparable to the frequency resolution of the experiment and therefore plays a small role. This explains why the CPT dip in Fig.\ 2 can be explained without taking into account charge noise-induced dephasing. The effects in Fig.\ 3(b) arise when $f_{\rm scan}$ is reduced even further. In this case, ultralow frequency flicker noise results in wanderings of the quantum dot optical resonance typically by a linewidth or two over the course of many minutes \cite{Kuhlmann2013}. These optical shifts induce in turn wanderings of the CPT dip, by a few MHz, as observed in Fig.\ 3(b). The larger shifts in CPT position, Fig.\ 3(b), may represent highly unlikely but more extreme changes to the electrostatic environment of the quantum dot.

Our results point to the role of charge noise in dephasing a quantum dot hole spin. In this particular experiment, the fluctuations in $E_{Z}^{h}$ imply a dephasing time of $T_{2}^{*} \simeq 100$ ns on integrating noise in a bandwidth starting around 1 Hz at $B=0.5$ T. Some quantum dots in the sample have considerably lower optical linewidths and for these we can expect $T_{2}^{*} \ge 100$ ns, consistent with CPT at lower resolution \cite{Brunner2009}. In fact, these resonant laser spectroscopy experiments have low charge noise, much less than in experiments with non-resonant excitation \cite{Kuhlmann2013}. Experiments with more charge noise will therefore give smaller hole spin $T_{2}^{*}$ values. In addition to a sample and method dependence, charge noise also results in a $B$-dependence of $T_{2}^{*}$. Our spectroscopy results, Fig.\ 4(a) and those at lower $B$, show that $dg_{h}/dF$ is $B$-independent, i.e.\ that the fluctuations in $E_{Z}^{h}$ increase linearly with increasing $B$ for constant charge noise. This implies that the dephasing induced by charge noise scales as $1/B$: 
\begin{equation}
T_{2}^{*} \simeq \frac{\hbar}{|dg_{h}/dF| \: \Delta F \: \mu_B B}.
\end{equation}
This $B$-dependence may be hard to detect if for instance other dephasing mechanisms come into play at higher magnetic fields, for instance hyperfine coupling \cite{DeGreve2011,Greilich2011} or the interaction with phonons \cite{Heiss2008,Trif2009}. We note however that hole spin dynamics at magnetic fields of several Tesla reveal smaller $T_{2}^{*}$ values \cite{DeGreve2011,Greilich2011} than those at low magnetic field \cite{Brunner2009,Crooker2010}, and this is consistent with charge noise-dominated spin dephasing. We stress that an advantage of the present experiment is that the charge noise is measured in situ via the laser spectroscopy.

In conclusion, we report 10 MHz wide CPT dips in laser spectroscopy experiments on a quantum dot hole spin. The quantum dot is embedded in a good but imperfect device. Imperfections in the device, notably charge noise, cause slow wanderings of the CPT dip. There are a number of mitigating strategies. First, p-type devices need to be developed with less charge noise, ideally with the low levels of charge noise associated with the best n-type devices. Secondly, the dependence of the hole g-factor on electric field, while possibly an attractive feature for electrical qubit control, can be reduced by appropriate quantum dot design. Thirdly, the small $\Delta R/R$ signals can be boosted to more practical levels by embedding the quantum dot in a resonant micro-cavity.

We acknowledge financial support from NCCR QSIT (RJW), Royal Society (BDG) and EPSRC (BDG).

\newpage

\end{document}